# System Design for the Event Horizon Imaging Experiment Using the PECMEO Concept

KUDRIASHOV Volodymyr    MARTIN-NEIRA Manuel    BARAT Itziar

MARTIN IGLESIAS Pertonilo    DAGANZO-EUSEBIO Elena

ALAGHA Nader    VALENTA Vaclav

(*European Space Research and Technology Centre, the European Space Agency, Noordwijk* 2201*AZ*)

**Abstract**    The concept for space interferometry from Polar or Equatorial Circular Medium Earth Orbits (the PECMEO concept) is a promising way to acquire the image of the "shadow" of the event horizon of Sagittarius A* with an angular resolution of circa 5 microarcseconds. The concept is intended to decrease the size of the main reflector of the instrument to about 3 m using a precise orbit reconstruction based on Global Navigation Satellite System (GNSS) navigation, inter-satellite range and range-rate measurements, and data from the Attitude and Orbit Determination System (AODS). The paper provides the current progress on the definition of the subsystems required for the concept on the basis of simulations, radio regulations, and available technology. The paper proposes the requirement for the localization of the phase centre of the main reflector. The paper provides information about the visibility of GNSS satellites and the needed accuracies of the AODS. The paper proposes the frequency plan for the instrument and its Inter-Satellite Links (ISLs). The concepts for measurement of range and range rate using ISLs (as well as for the data exchange at these ISLs) are presented. The block diagram of the interferometer is described and its sensitivity is estimated. The link budget for both ISLs is given as well as their critical components. The calculated measurement quality factors are given. The paper shows the expected performance of the sub-systems of the interferometer. The requirements for the localization of the main reflectors and the information about the availability of the GNSS satellites are based on the simulations results. The frequency plan is obtained according to the PECMEO concept and taking into account the radio regulations. The existing technology defines the accuracies of the AODS as well as the link budgets and the measurement accuracies for both ISLs and the sensitivity of the instrument. The paper provides input information for the development of the orbit reconstruction filter and the whole PECMEO system.

**Classified index**    P 412

## 0    Introduction

The proximity of Sagittarius A* (Sgr A*) makes the characteristic angular size scale of its Schwarzschild radius larger than for any other candidate black hole namely 10 microarcseconds ($\mu$as) which corresponds to the observable diameter of 50 $\mu$as[1].

All existing interferometers with very long baselines involve a ground segment and use fringe detection to carry out their observations. Atmosphere variation limits the integration time and affects the phase of both incoming and downlink signals of the interferometers. The aperture size which can be synthesized on the ground is limited by the size of the Earth itself. Space to Earth interferometry allows extremely large baselines, but the observation wavelength is limited by the satellite localization accuracy.

The considered concept for space-to-space interferometry[2] is dedicated to mitigating the requirement (of about $10\,\mathrm{m}^{[3-5]}$) to the antenna size by means of the employment of GNSS-based navigation and inter-satellite measurements. As well, the novelty of the concept consists of the absence of any ground telescope (aimed to withdraw the atmospheric phase error), the usage of frequencies more than 200 GHz (aimed to reduce image broadening due to the interstellar scattering), the occupation of polar orbits with small difference in orbit radii (aimed to benefit from the spiral $uv$-coverage[6]), and the employment of twin satellites (aimed to optimize both the efficiency and the cost).

The paper is devoted to propose the localization requirement for the phase centre of the main reflector, to provide the GNSS visibility for the selected orbit configuration, to show the performance estimates of Inter-Satellite Links (ISLs) dedicated for both communication and satellite localization tasks, and to present the sensitivity calculation.

# 1 Orbit Selection and the Longest Baseline

The intrinsic diameter of Sgr A* observable at a wavelength of 1.3 mm is $37^{+16}_{-10}\,\mu\mathrm{as}^{[1]}$. The aimed angular resolution of the imaging instrument is $5\,\mu\mathrm{as}^{[2]}$. The angular resolution is limited by $\lambda/D$ ($\lambda$ denotes the wavelength and $D$ is the maximal aperture). At a perfectly efficient projected aperture, the aimed ratio $\lambda/D$ is of $2.4\times10^{-11}$.

On the one hand, the interstellar medium disables the generation of a detailed image of Sgr A* for wavelengths longer than 0.6 mm[7]. On another hand, the antenna element localization (knowledge) error limits the synthesized aperture combining efficiency[8]. The selected input wavelength band corresponds to a sub-mm requirement in the accuracy of the interferometer baseline localization.

The highly elliptical orbit of the Spektr-R satellite (RadioAstron Project) is mostly above the GNSS constellations[3]. The orbit apogee is of circa $340\times10^3$ km which can be used for synthesizing an aperture that large. The orbit determination accuracy of the Spektr-R reached several tens of meters and millimetres per second in 3-Dimensional (3D) position and velocity, respectively[9]. This orbit determination accuracy disables operation at sub-mm wavelengths in such orbit altitudes above GNSS.

In Low Earth Orbit (LEO), the accuracy of the satellite orbit determination (of circa 3 cm) is mainly limited by the presence of the rapidly-changing ionospheric delays[10,11]. The simulated accuracy of relative navigation in LEO is down to millimetre level[12].

At near-circular, polar orbit, a navigation accuracy down to 1.5 mm and $5\,\mu\mathrm{m\cdot s^{-1}}$ for relative position and velocity, respectively have been simulated, at an orbit altitude of 450 km, i.e. even within the effect of the ionospheric errors[13]. Further work on the considered PECMEO concept[2] may consider the addition of input data channels for an extended Kalman filter and the post-processing of cross-correlation functions of the instrument signals in order to achieve the baseline localization accuracy knowledge of circa 0.1 mm.

A twin spacecraft space VLBI system has been proposed at circular orbits with a maximum altitude above the GNSS orbits[6]. The concept provides a densely populated $uv$-coverage. In contrast, the PECMEO concept mitigates the requirement on the sensitivity of each baseline ($uv$-point) by employing GNSS navigation, and inter-satellite measurements[2].

The PECMEO concept may be promising for operating with the main reflector diameter about 3 m instead of about 10 m diameter of Spektr-R, Spektr-M, and HALCA[3-5]. Additionally, the PECMEO concept provides connected element interferometry thanks to the generation of the local oscillator from a mix of the clocks of the twin satellites[14].

## 1.1 Orbit Orientation and Maximum Orbit Radius

The circular polar orbit has been selected in the PECMEO concept paper[2]. A 90° inclined polar or-bit has the property to maintain constant its orbital plane in the inertial reference system[2]. Therefore, the angle between the orbital plane and the centre of the galaxy remains constant. The difference in radii of the 2 above-mentioned orbits will drive the speed of change of inter-satellite distance (the interferometer baseline). The eccentricity vector of the orbit will drive the baseline variations within an orbit. An ideal circular orbit will result in the monotonous baseline change. Thus, the inclination of the proposed orbit is 90°. To mitigate the aperture projection loss towards the Sgr A*, the longitude of the ascending node of this orbit is 176.4°.

Operation outside the Van Allen radiation belts is desired[15]. Such orbit radii are of circa 12 400~19 100 km, and over 25 800 km[16]. Albeit, these belts are not stable in time and their day and night sides are not symmetric between themselves due to the influence of the solar wind[17−19]. The angular resolution depends on the effective aperture size and hence, the highest possible orbit is preferred.

Precise carrier phase (ambiguity solving) navigation requires at least 4 GNSS satellites to be in the common visibility of both Event Horizon Imaging Experiment (EHIE) satellites, even when they are furthest apart. The inter-satellite ranging is devoted to mitigating the requirement to 3 GNSS satellites. To assure the mitigation, the ranging signal must provide the best ranging accuracy and precision. It is initially assumed that 2 navigation antennas are mounted on opposite sides of each EHIE satellite. The orbit radii allow avoiding the ionosphere and hence, carrier-phase navigation can be done at a single carrier frequency only. However dual-frequency observations are preferred since they provide more independent measurements, which is needed to achieve the tight relative navigation requirements.

Considering 4 GNSS constellations available, the numbers of GNSS satellites in the common Field of View (FoV) are 5 and 2.3 at the orbit radii 10 152 and 15 229 km, respectively. A linear interpolation allows calculating the maximal orbit radius of 13 913 km. The selected radius provides, on the one hand, the aimed as-long-as-possible baselines and, on the other hand, an orbit below GPS and Galileo, and between the Van Allen belts. The mentioned below update of the navigation antennas to the single dipole allows additional flexibility in the maximum orbit radius.

### 1.2 Longest Baseline

For transfer of telemetry and for additional measurements of the baseline $D$ and its rate of change, an ISL at K-band is selected (Section 3). At this frequency band, the one-way attenuation is less than $0.1\,\text{dB}\cdot\text{km}^{-1}$ for clear atmosphere and of circa $0.5\,\text{dB}\cdot\text{km}^{-1}$ for $5\,\text{mm}\cdot\text{h}^{-1}$ rain rate[20].

According to the PECMEO concept, both the clock signals and the observation signals (receivers output signals to be cross-correlated) are exchanged between satellites[2]. The V-band is selected for these ISLs (Section 3) due to the selected (Medium Earth) orbit, the impact of the signal bandwidth on the sensitivity, and the available technology.

The occultation (or blockage) of the V-band ISLs disables any operation. At 60 GHz frequency, the one-way attenuation is more than $10\,\text{dB}\cdot\text{km}^{-1}$ and less than $3\,\text{dB}\cdot\text{km}^{-1}$ for clear atmosphere and $5\,\text{mm}\cdot\text{h}^{-1}$ rain, respectively (less $0.5\,\text{dB}\cdot\text{km}^{-1}$ at weaker rain and fog) because the oxygen absorption peak is observed at circa 60 GHz that establishes the high clear atmosphere absorption[20]. The blockage of the signal by the Earth atmosphere and ionosphere limits the longest baseline of the interferometer as follows:

$$D = \sqrt{r_1^2 - (r_\text{E} + r_\text{T})^2} + \sqrt{r_2^2 - (r_\text{E} + r_\text{T})^2}, \quad (1)$$

where $r_1$ and $r_2$ denote orbit radii (to be valued below) of both satellites, $r_\text{E} = 6371\,\text{km}$ is the Earth radius, and $r_\text{T}$ is the thickness of both troposphere and ionosphere that is assumed to equal to 20 km, at the occultation of the ISLs.

The paper considers 4 options for the longest baseline $D$ in Eq. (1), at the highest orbit radius $r_2 \approx 13\,913\,\text{km}$ (Table 1). The shortest baseline is equal to $d = r_2 - r_1$.

The Sun radiation captured by the observation antennas (main reflectors) produces a bias. If a thermal shield layer protected the back side of the main reflector, then the longest observation time would be the 6-month period when the Sun is in the back of the orbital plane. The corresponding longest baseline

is 24 712 km (Table 1).

The expected Sgr A* variability time scale (of minutes) is much shorter than observing time scale (of months) at the realistic technical performance (Subsection 4.2) and hence, the time-averaged image only can be reconstructed. Exploring time-dependent source models may widen understanding of the origin of flaring structures in Sgr A* and improve the capabilities of the imaging techniques[21]. The quiescent source morphology can be obtained by observing over multiple time-epochs and subsequent processing of the visibilities (scaling, averaging, and smoothing) before imaging[21].

## 2 Relative Localization of the Interferometer

The continuous sending of the observation signals to the ground, in order to cross-correlate them, is impossible due to both the atmospheric blockage and the narrow allocated bandwidth for a radio-frequency downlink[22]. Therefore, the cross-correlation must be done in orbit[2]. The geometric delay of the incoming signals (to be compensated in the cross-correlator) is determined within the GNSS navigation uncertainty and hence, the signals must be cross-correlated in the corresponding number of delays. The data rate required to downlink the whole number of cross-correlation functions is much less than that needed to downlink the measured signals. The cross-correlations are to be sent to the ground. An extended Kalman filter is dedicated to the orbit reconstruction.

The main differences of the navigation in the PECMEO concept and in the above-mentioned (formation-flying) LEO systems are that the orbit radii alleviate the ionospheric delay. The longest inter-satellite distance is larger and the geometry is different. At the output of the filter, the localization accuracy is better than the accuracy of the above-mentioned GNSS navigation. The latter allows reducing the cross-correlation data package. A technique which allows finding the true cross-correlations in the data package has not been established yet. The final image reconstruction is to be performed on the ground.

### 2.1 Relative Localization Knowledge Requirement

The observation wavelength $\lambda$ shorter 0.6 mm is required for the aimed angular resolution, at the available $D$ (Table 1). The accuracy of the relative localization (knowledge) error of the phase centre between the receiving antennas constrains the combining efficiency and hence the directivity of the antenna being synthesized, assuming that the baseline changes between measurements only[8].

The directivity of the synthesized antenna has been simulated with respect to the Standard Deviation (STD) of the localization knowledge error of the array elements. At each value of the STD, 50 realizations of all 3 Cartesian coordinates of all positions of the antennas along the orbits were realized using (not-correlated between themselves) random errors, as the first step. Point spread functions were generated, as the next step. Further, they were averaged. The averaged point spread function was employed to calculate the directivity in the direction of Sgr A* as the ratio of the normalized radiation pattern of the antenna to the average value of the pattern[23]. The standard interferometry equations, as well as the D'Addario equation, are not given for brevity. The simulation parameters are as follows.

Table 1  Shortest and the longest baseline lengths, and duration of observation period as a function of the lower orbit radius $r_1$, assuming the higher orbit radius of $r_2 \approx 13\,913$ km

| Orbit radius $r_1$/km | 13 892 | 13 903 | 13 906 | 13 909 |
| --- | --- | --- | --- | --- |
| Shortest baseline $d$/km | 20.35 | 9.7 | 6.5 | 3.25 |
| Longest baseline $D$/km | 24 693 | 24 705 | 24 709 | 24 712 |
| Observation time duration/month | 1 | 2 | 3 | 6 |

(i) The emission source is point-like, the distance to the source is $5\times10^9$ m (0.033 AU), the azimuth of the source is $0°$, the source elevation is $-29°$ and geometric delays are calculated using the Euclidean distances.

(ii) The incoming (analytic) signal is a single tone.

(iii) The errors (in knowledge) of the locations of the phase centres are not correlated between all 3 spatial coordinates and hence, $\sigma_{3D} = \sqrt{3}\sigma$, where $\sigma = \sigma_x = \sigma_y = \sigma_z$ denotes the std which varies providing $\sigma_{3D}$ up to 1/3 of centre wavelength (Fig. 1).

The simulation mitigates the localization requirement circa 20% compared with the D'Addario equation (Fig. 1). The result has been approximated by the modified D'Addario equation in which the index of the exponent is multiplied by $4^{-1/3}$. The accepted loss of area of the synthesized antenna due to the localization (knowledge) error is not more than 2 times. It is assumed that the adopted loss corresponds to a loss of 3 dB in antenna directivity[23]. The simulation shows that the 3 dB directivity loss occurs at $\sigma_{3D} = \lambda/6$ (Fig. 1).

The required $\sigma_{3D}$ is below circa $80\,\mu$m ($\sigma \approx 45\,\mu$m), at the highest frequency 637 GHz of the incoming signal (see Subsection 3.4). The required $\sigma_{3D}$ are of 158 and $223\,\mu$m for the centre frequencies of 323 and 229 GHz, respectively. The baseline localization knowledge accuracies $\sqrt{2}\,\sigma_{3D}$ aimed at 229 and 637 GHz are of 0.32 and 0.11 mm, respectively.

## 2.2 GNSS Signals Availability and Navigation Antenna

The visibility of GPS satellites has been simulated in the Systems Tool Kit software. The simulation considered that the pointing of each GPS satellite antenna is towards the centre of the Earth and the beam width of the GPS antenna is $40°$. In the simulation, the semi-major axes of the EHIE satellites are 13 898 and 13 918 km, i.e. the 1-month observation time (similarly to Table 1). The number of GPS satellites visible from each EHIE satellite is up to 10. The number is similar for both EHIE satellites due to the small difference in semi-major axes of their orbits. Each navigation signal is received during 2-time intervals: (i) when the EHIE satellite is between the GPS satellite and the Earth and (ii) when the EHIE satellite is past the Earth between the Earth blockage and the GPS antenna pattern blockage[2].

As the baseline length extends during the observation period, the common visibility of the navigation antennas reduces. That is why the number of the GPS satellites in the common visibility is inverse to the operation time (Fig. 2). The mean time duration of GPS satellites in common visibility is of circa 7.1∼22.1 min. The maximum time in common visibility is 21.2∼153.5 min.

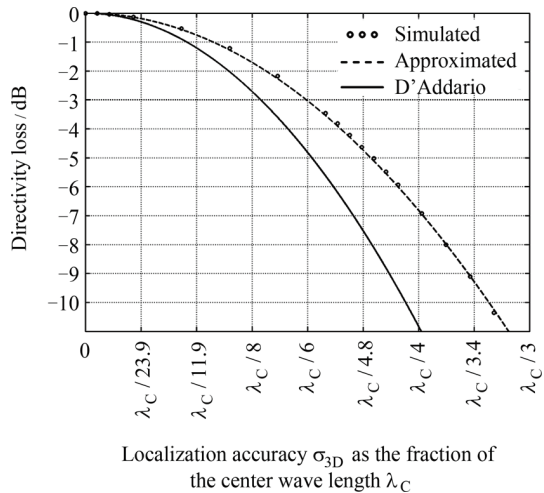

Fig. 1　Directivity loss of the synthesized antenna array with respect to the STD of the localization knowledge error of phase centres of the array elements, given as the fraction of the center wavelength

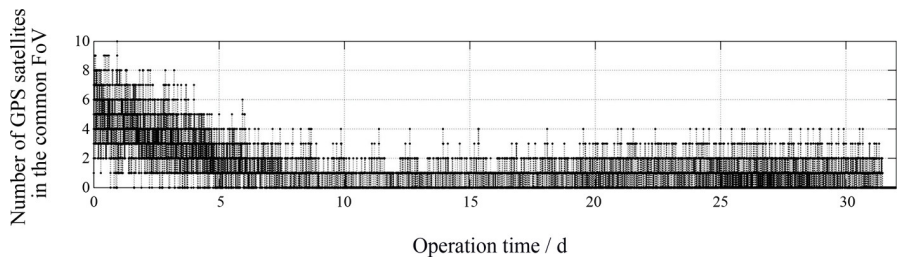

Fig. 2　Number of GPS satellites in the common Field of View (FoV) with respect to the operation time

The number of GPS satellites in the common visibility is up to 10, at the shortest baseline lengths only (Fig. 2). There are time intervals of unavailability of GPS satellites in the common visibility. Operation during these intervals could be supported by data from GNSS satellites in individual visibility, other sensors and an extended Kalman filter.

The consideration of all GNSS constellations instead of the GPS only is expected to increase the visibility of GNSS satellites 4 times[24].

Histograms of directions of arrival of GPS signals to the EHIE satellites allowed to establish the requirements to the navigation antenna pattern. The pattern must be omnidirectional in azimuth. The antenna beamwidth in elevation must be 90°. The preferred navigation antenna is the half-wavelength dipole which is mounted to the satellite on a dielectric stick. The null of the toroidal-shaped antenna pattern is pointed along the normal to the orbit. The peak gain, of 2.15 dB, is achieved for any direction within the orbital plane. The dipole has a loss of 3 dB due to its operation at a single polarization.

### 2.3 Expected Quality of the Relative Localization

A transfer of the phase centre of the main reflector to the phase centres of the navigation and the ISLs antennas is not perfect. Mounting biases of the spacecraft structure have to be characterized on the ground through optical means using cubic mirrors (a standard procedure). The orbit maneuver which allows equal radii of the EHIE satellites orbits is dedicated to measure the bias which remains after the launch[2]. The positioning (orientation) system is planned to be based on an extended Kalman filter. Orbital parameters of EHIE satellites are to be continuously determined using GNSS measurements, inter-satellite range-rate measurements from V-band ISLs, inter-satellite range measurements from K-band ISLs, star trackers, accelerometers, gyroscopes, equations of motion, direction of arrival of an external calibration source signal, and optional ground-based lidars.

Due to the absence of ionospheric effects at the chosen orbit radii, the expected real-time "on flight" GNSS-based relative navigation accuracy $\sigma_{3DN}$ is better than 5 cm[2]. The accuracies of range and range-rate measurements given in Sections 5 and 6 are of circa 30 $\mu$m and below 0.1 $\mu$m·s$^{-1}$, respectively. Both the latter values allow solving the relative navigation ambiguity and measuring the baseline length (and its rate), not its orientation. The accuracy (1-sigma) of a high performing star tracker is better 1 arcsec[25]. The accuracy of accelerometers developed for the GOCE mission is from $10^{-10}$ to $10^{-12}$ m·s$^{-2}$ [26]. The accuracy of a gyroscope is better 1 arcsec ①. The laser ranging with the ground-based lidar may provide an accuracy of 0.1 m[3]. At the output of the filter, the expected baseline localization knowledge accuracy is the aforementioned $\sqrt{2} \times \sigma_{3D}$ value.

## 3 Frequency Plan

Several key aspects have been taken into account in the selection of the frequency bands for the operation of this passive instrument. Both the interstellar scattering and the Sgr A* emission behavior enforce aiming observation frequencies well above 200 GHz with the preferable option of circa 600 GHz. The observations of the passive instrument are to be performed in the bands allocated by the ITU Radio-Regulations to the Space Research Service (passive). Some of the identified candidate frequency bands are 226∼231.5, 313∼356, 634∼654, and 657∼692 GHz[22]. The front end of the instrument relies on the technology of the MetOp-SG and Juice SWI②. The input frequency bands of the instrument could overlay observation frequencies of the Event Horizon Telescope at circa 230 and 345 GHz[2]. The highest frequency is limited by the directivity loss (Fig. 1), this being 634∼654 GHz band.

The observation signals are exchanged between satellites through the ISLs in order to be cross-correlated on the fly. Two permanent simplex ISLs

---

① https://spaceequipment.airbusdefenceandspace.com/avionics/fiber-optic-gyroscopes/astrix-200/

② https://directory.eoportal.org/web/eoportal/satellite-missions/m/metop-sg

are dedicated to exchanging both observation signals and the ISL carrier frequency signals between the satellites. The attainable bandwidth $B_{\max}$ of the instrument limits the sensitivity as $B_{\max}^{-0.5}$ and hence, the widest option is sought[27,34]. The desired bandwidth of more than 1 GHz presents a challenge for currently available digital technology[①] and hence, analogue transmission is considered for the ISLs. Radio frequency technology for ISLs allows frequency bands up to 100 GHz. The range of the V-band 59.3~71 GHz is the widest suitable allocation for ISLs between non-geostationary satellites[22].

Both the sum and the difference of the frequencies of the carrier signals of the V-band ISLs are needed to obtain the frequencies of local oscillators dedicated to the down-conversion of the observation signals[2]. All frequencies are to be derived from a 10 MHz master clock by frequency multiplication and amplification.

The inter-satellite ranging is dedicated to assisting in solving the integer ambiguities of GNSS carrier phase observables provided on wavelengths of circa 0.19 and 0.25 m. For this purpose, the ranging signal must assure the best achievable range resolution and accuracy, *i.e.* the ranging is aimed at the widest input frequency bandwidth, the highest signal-to-noise ratio (SNR), and the lowest interference to the instrument observation bands. Utilizing both the exchange of the observation $B_{\max}$ and ranging within the same V-band is challenging. Therefore, the ranging is to be done outside the V-band frequency allocation, which is then only dedicated to exchange observations and oscillator carriers. The next widest allocation band is at frequencies 25.25~27.5 GHz[22] called K-band in the paper. The K-band allocation is used for ranging, as in the GRACE mission[②]. In addition, this frequency band is used for transferring the relative navigation telemetry and as the redundant range-rate measuring system.

### 3.1 V-band ISLs for the Exchange of the Instrument Signals

The V-band ISL is based on single sideband amplitude modulation. The frequency plan of the V-band ISLs (Fig. 3) is calculated from the following expressions:

$$\begin{cases} f_1 = f_{\text{RR,min}} + c_0, \\ B_{\max} = \dfrac{f_{\text{RR,max}} - 2c_1 - (f_1 + c_1)(w-1)}{w}, \\ f_2 = (f_1 + c_1 + B_{\max})(w-1), \end{cases} \quad (2)$$

where $f_{\text{RR,min}} = 59.3$ GHz and $f_{\text{RR,max}} = 71$ GHz denote frequencies of the ISL allocation band[22], $f_1$ and $f_2$ denote carrier frequencies of these ISLs, $B_{\max}$ denotes the widest attainable band width to be exchanged between satellites, $c_0 = 0.1$ GHz denotes a frequency margin between $f_{\text{RR,min}}$ and $f_1$, $c_1 = 0.1$ GHz denotes the frequency margin between the LO (either $f_1$ or $f_2$) and the spectrum as in the MetOp-SG, $w = 2\left(1 - \dfrac{w_{\text{BPFs}}}{2}\right)^{-1}$ and $w_{\text{BPFs}} = 3/100$ denote the fractional bandwidth of the separation between the two band pass filters that assures the isolation (defined below) between the transmit and receive bands of the ISLs.

Filters at both receiver and transmitter ends assure the isolation of circa 100 dB between two non-overlapping frequency bands. The fractional bandwidth of the gap between the bands must be of about 3% in order to keep widest possible radiometer band-

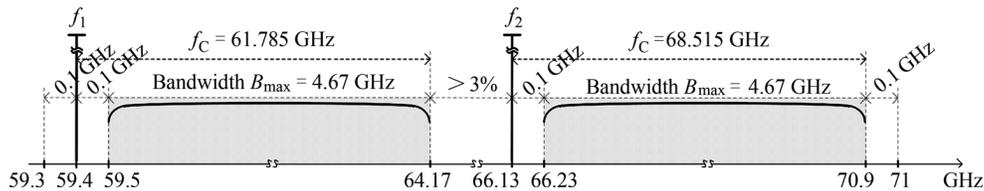

Fig. 3   Frequency plan for the V-band inter-satellite links

---

[①] https://www.iis.fraunhofer.de/content/dam/iis/en/doc/ks/hfs/Flyer_Solutions%20for%20DVB-S2X%20Wideband%20Transmission.pdf

[②] http://www2.csr.utexas.edu/grace/

width and account for the filter insertion loss. A numerical simulation confirmed that such filter performance is realistic. The insertion loss of such filters is above 1 dB.

Both $f_1$ and $f_2$ must be derived from the master clock and hence, the V-band ISLs frequency Plan 2 is as follows (in GHz): $f_1 = 59.4$, $f_2 = 66.13$, $B_{\max} = 4.67$.

## 3.2 Local Oscillators

The frequencies of both the RF and IF (Intermediate Frequency) local oscillators as well as the input frequency bands of the instrument are derived from the V-band ISLs carrier frequencies $f_1$ and $f_2$ as follows:

$$\begin{cases} f_{\text{LO,RF}} = c_{\text{m}}[f_1 + f_2 - (f_1 + f_2)/c_{\text{LO,RF}}], \\ f_{\text{LO,IF}} = f_2 - f_1, \\ f_{\text{In,min}} = f_{\text{LO,RF}} + f_{\text{LO,IF}} + c_1, \\ f_{\text{In,max}} = f_{\text{In,min}} + B_{\max}, \end{cases} \quad (3)$$

where $f_{\text{LO,RF}}$ denotes the frequency of the LO for the frequency down conversion from RF to the first IF, $c_{\text{m}}$ denotes the coefficient of the frequency multiplication, $c_{\text{LO,RF}}$ is the frequency division coefficient addressed below, $f_{\text{LO,IF}}$ denotes the frequency of the LO for the frequency down-conversion from the 1st IF to the 2nd IF, called here the baseband (the shift between the DC and the 2nd IF is $c_1 = 0.1\,\text{GHz}$, and $f_{\text{In,min}}$ and $f_{\text{In,max}}$ denote the lower and upper frequencies of the input observation bandwidth $B_{\max}$, respectively. $B_{\max}$ is limited by Eq. (2) and by the number of input frequency channels.

Adjustment of the frequency division coefficient $c_{\text{LO,RF}}$ allows (i) simplifying the frequency multiplication scheme, (ii) matching the allocated/identified observation bands, (iii) satisfying requirements to both the angular resolution and the localization knowledge, (iv) optional matching the EHT and/or MetOp-SG bands, and/or molecular lines observable towards the Sgr A*, keeping the symmetry of the concept[1,2,22,28,29]. Accordingly, the $c_{\text{RO,LF}}$ of 8 is selected for the lowest frequency band only while both higher frequency bands ($c_{\text{m}} = 3$ and $c_{\text{m}} = 6$) employ $c_{\text{LO,RF}} = 6$.

The first local oscillator frequencies $f_{\text{LO,RF}}$ calculated for $c_{\text{m}}$ of 2, 3 and 6 are 219.6, 313.8, and 627.7 GHz, respectively. The 2nd local oscillator frequency is $f_{\text{LO,IF}} = 6.73\,\text{GHz}$.

## 3.3 Science Instrument Observation Bandwidth

Taking more than 1 observation channel decreases the bandwidth $B_{\max}$ in Eq. (2) i.e. degrades the instrument sensitivity[27]. It is proposed to have a band-pass filter at the output of a latch following a 1-bit Analogue-to-Digital Converter (ADC) which allows retrieving harmonics of the baseband signal at outputs of dedicated band-pass filters (Fig. 4)[29]. The sampling frequency and the attainable maximum bandwidths are as follows:

$$\begin{cases} B_2 = \dfrac{B_{\max} - 2c_1(c_A - 1)}{2c_A}, \\ B_3 = \dfrac{B_{\max} - 2c_1 c_A}{2c_A + 1}, \\ f_{\text{S},i} = 2c_A(c_1 + B_i). \end{cases} \quad (4)$$

where $B_i$ are the frequency bandwidths for $i = 1, 2$, channels, $f_{\text{s}}$ is the sampling frequency which is $c_A$ times higher the Nyquist frequency.

Accounting for aliasing and the sensitivity requirement, the $c_A = 1.05$ assigns filter requirements of the instrument stricter than the $w_{\text{BPFs}} = 3/100$. The $c_A$ value allows calculating the following bandwidths: $B_1 = B_{\max} = 4.67\,\text{GHz}$, $B_2 = 2.22\,\text{GHz}$ and $B_3 = 1.44\,\text{GHz}$.

Thus, the concept of over-sampling the latched ADC output allows additionally the above-mentioned tight arrangement of the frequency bands $B_i$ within the bandwidth $B_{\max}$.

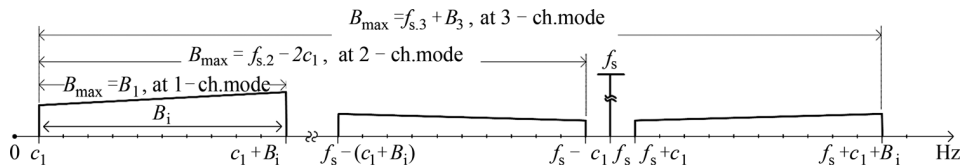

Fig. 4 Science instrument observation bandwidth with respect to the number of observation bands

### 3.4　Science Instrument Observation Bands

The lowest input frequencies in Eq. (3) for $c_m = 2$, 3, and 6 are of circa 226.5, 320.7 and 634.5 GHz, respectively. The bandwidth options $B_i$ in Eq. (4) allow calculating the highest input frequencies using Eq. (3). At the widest option $B_{max} = 4.7$ GHz, the single input frequency channel is available either at 226.5∼231.2 GHz, 320.7∼325.3 GHz, or 634.5∼639.2 GHz (Table 2). Narrowing the input bandwidth to $B_2 \approx 2.2$ GHz allows the simultaneous operation at 2 input frequency bands being selected from 226.5∼228.7, 320.7∼322.9, and 634.5∼636.7 GHz. The narrowest input bandwidth $B_3 \approx 1.4$ GHz allows the simultaneous operation at all 3 input frequency bands of circa 226.5∼228, 320.7∼322.1, and 634.5∼636 GHz (Table 2).

Molecular resonant peaks were measured towards the Sgr A*[28]. Molecular emission is typically resolved by a VLBI. Moreover, those peaks do not overlay the candidate observation frequency plan.

### 3.5　K-band ISLs

The K-band allocation is shared by both the dual frequency one-way ranging (and range-rate) signals and the telemetry signal which allows exchanging GNSS information between the satellites.

The relative navigation is the critical aspect of the concept and hence, the necessary telemetry information may be embedded into the ranging code as recommended by the Consultative Committee for Space Data Systems. The pseudoranges and carrier phase observations from the chosen set of GNSS satellites at every time epoch are to be exchanged between the 2 satellites. The transmission is asynchronous, happening typically at 1 Hz, and the data rate of the telemetry signal is of circa 2 kbit·s$^{-1}$. The current work considers the amplitude modulation of the telemetry signal but other types of modulation are possible. The frequency plan of the K-band ISL is calculated as:

$$\begin{cases} f_{K,1} = f_{K,RR,min} + c_1/2, \\ B_{K,max} = \dfrac{2.25 - 3c_1 - w_K(f_{K,RR,min} + 1.5c_1)}{2 + w_K}, \\ f_{K,2} = (f_1 + c_1 + B_{K,max})(1+w), \end{cases} \quad (5)$$

where $f_{K,RR,min} = 25.25$ GHz denotes the lowest frequency of the ISL band allocation[22], $f_{K,1}$ and $f_{K,2}$ denote the frequencies of the local oscillators of these ISLs, $B_{K,max}$ denotes the widest attainable bandwidth that can be exchanged between the satellites, $c_1/2$ is the frequency margin between the allocation band edges of these ISL bands, the $c_1 = 0.1$ GHz is the frequency margin between the LO and the spectrum, $w_K = w_{K,BFPs}\left(1 - \dfrac{w_{K,BFPs}}{2}\right)^{-1}$ and $w_{K,BPFs} = w_{BPFs} = 3/100$ denote the fractional bandwidth between two band-pass filters that assures the isolation defined below between ISLs.

Both $f_{K,1}$ and $f_{K,2}$ must be multiples of the local master clock. The calculated $f_{K,2}$ is 26.7706 GHz. Aiming to keep the frequency margin below 27.5 GHz, the bandwidth is to be slightly decreased beyond Eq. (5). With $f_{K,2} = 26.77$ GHz, the widest possible bandwidth is expressed as follows:

$$B_{K,max} = \dfrac{f_{K,2}}{1+w} - (f_1 + c_1) = 578.8 \text{ MHz}. \quad (6)$$

The frequency plan of the K-band ISLs is calculated from Eq. (5) and (6) those yield to: $f_{K,1} = 25.3$ GHz, $f_{K,2} = 26.8$ GHz, and $B_{K,max} = 578.8$ MHz.

## 4　Science Instrument

### 4.1　Instrument Architecture

The main reflector performance assumptions rely on ESA's Planck carbon fiber reinforced plastic honeycomb sandwich technology[30]. The desired parame-ters of the main reflector antenna are as follows: 3 m projected reflector diameter, 8.6 μm reflector RMS surface error, 0.96 antenna radiation efficiency (Ohmic loss of the antenna material), more than 0.65 antenna aperture taper efficiency and 0.99 antenna spillover efficiency[30]. In our preliminary concept, the main reflector is deployed around a 1-axis hinge mounted onto the satellite.

**Table 2　Input frequency bands for 3 observation bandwidths options**

| $c_m$ | $f_{In,min}$/GHz | $f_{In,max}$/GHz | | |
|---|---|---|---|---|
| | | $B_{max}$ | $B_2$ | $B_3$ |
| 2 | 226.5 | 231.2 | 228.7 | 278 |
| 3 | 320.7 | 325.3 | 322.9 | 322.1 |
| 6 | 634.5 | 639.2 | 636.7 | 635.9 |

Frequency selective surfaces allow separation of input frequency bands. Polarization selective surfaces allow splitting the incoming radiation, as to be fed into H and V horns. Cooled down receiver follows each horn.

The desired noise temperature of Low Noise Amplifier (LNA) is less than 480 K, at the highest input frequency band (Table 3). At a 600 GHz band, the best known LNA offers a noise temperature below 2400 K that can be improved to 600 K, at a physical temperature of 24 K[31−33]. Thus, the development of the cryo-cooled LNA is needed.

The power consumption of the cryocooler depends on both the temperature and the power to be dissipated by the cooler. The expected power to be dissipated is of circa 0.66 W, accounting all 6 input frequency LNAs and mixers. The latter physical temperature of 24 K is challenging. The power consumption of the cryocooler is 330 W, at the adopted temperature of 55 K.

This space-to-space interferometry concept[2] involves 1 to 3 input frequency channels (Table 2). At any channel, the frequency of the incoming radiation is down-converted twice[2]. Both local oscillators $f_{\text{LO,RF}}$ and $f_{\text{LO,IF}}$ dedicated to that purpose are derived from the mixing of the carrier frequencies of the ISLs. These are exchanged between the satellites to achieve the connected-element mode. At each satellite, the received frequency is Doppler shifted by less than $\pm 0.9$ kHz due to the same inter-satellite velocity, at this difference between carrier frequencies. Both carrier frequencies of the ISLs are multiples of the master clock frequency and the frequency conversion details are given in the concept paper[2].

The frequencies of the local oscillators are derived from $f_1$ and $f_2$ (2) according to Eq. (3). The output signals of the RF mixers and fed into the IF amplifier. The IF signal is fed into the band-pass filter. The bandwidth of the filter $B_i$ is derived from Eq. (4) according to subsection 3.3. The output signal of the filter is down converted to the $c_1 - B_i$ frequency band. This baseband signal is then fed into 1-bit ADC with the sampling frequency $f_{\text{S},i}$ (4). The output signals of the ADC are fed to both the cross-correlator whose output is time-stamped, stored and eventually sent to the ground and to the V-band ISL devoted to send the measured instrument signals to the other satellite. Aimed to improve the sensitivity output signals are combined together to acquire the Stokes I parameter of the incoming radiation.

Table 3  System parameters and the sensitivity at 637 GHz

| Parameter | Value | Reference |
|---|---|---|
| Antenna temperature $T_{\text{A}}$/K | 9.35 | Subsection 4.2.1 |
| Receiver temperature $T_{\text{R}}$/K | 504 | Subsection 4.2.1 |
| 1-bit sampling efficiency $\frac{1}{\sqrt{Q}}$ | 0.63 | Subsection 4.2 |
| Input bandwidth $B_2$/GHz | 2.22 | Eq. (4) |
| Observation time $\tau_{\text{obs}}$/month | 6 | Table 1 |
| Main reflector diameter/m | 3 | Subsection 4.1 |
| Effective area/m$^2$ | 4.14 | Subsection 4.2.2 |
| Longest Baseline $D$/km | 24 712 | Table 1 |
| Effective synthesized area $A_{\text{syn,e}}$/m$^2$ | $0.21 \times 10^{15}$ | Subsection 4.2.2 |
| Sensitivity $\Delta T$/K | $155 \times 10^6$ | Eq. (7) |
| Standard deviation of the additive thermal noise $\sigma_{\text{noise,e}}$/mJy | 1.02 | Eq. (8) |
| Source flux in "halo"/mJy | 5 | Subsection 4.2.3 |
| Signal-to-noise ratio in "halo" | 4.9 | Subsection 4.2.3 |

The correlator delay lines depth is $\pm 2\sigma_{3DN} = \pm 10$ cm[2]. The longest delay to be compensated is $D(1 + \sin\alpha) \approx 36\,695.8$ km, where the angle $\alpha = 29°27.9'$ is the declination of Sgr A*, and $D = 24\,712$ km is the longest inter satellite distance (Table 1) and hence, the longest $\tau$ is of circa 0.12 s.

The on flight cross-correlation is performed in 3 steps[2]. First, one of the signals $s_1$ goes through a delay line which includes an interpolator circuit to realize any continuous value of delay. The delayed signal $s_1$ enters then a tapped delay line, each tap output being multiplied with the second signal $s_2$. The next step consists in the removal of the Doppler frequency shifts of both local oscillator component and incoming signals using the phase models given in Ref. [2]. The last step is the accumulation of the phase-corrected products. In order to be sent on the ground, both the accumulated and the set of delays are timestamped and stored into memory. The latter is done for each integration time. The time-stamped ISLs measurements are stored in the memory too.

The refinement of the delay is to be performed on the ground in 2 further steps. The orbit reconstruction filter allows narrowing the set of the delays, as the first step. The refinement of this set to the required localization accuracy ($\sqrt{2}\sigma_{3D} \approx 0.1$ mm) is then done, as the second step. Employment of a lower-frequency channel is addition to the localization concept (Subsection 2.3) is not described in this paper.

### 4.2　Sensitivity Calculation

The sensitivity is calculated as follows[27]:

$$\Delta T = \frac{T_A + T_R}{\sqrt{2B_i\tau_{obs}}} \frac{A_{syn,e}}{A_{real,e}} \sqrt{Q}, \qquad (7)$$

where $T_A$ and $T_R$ denote the antenna and the receiver noise temperatures, respectively, $B_i$ is the input frequency bandwidth, $\tau_{obs} = 6$ months denotes the overall observation time that is the sum of integration time of all $uv$-points ($\tau_{obs}$ is limited by the time when the Sun is in the back of orbital plane), $A_{syn,e}$ and $A_{real,e}$ denote the effective area of the synthesized aperture and the main reflector, respectively. The factor $Q = (\pi/2)^2 \approx 2.46$ is the sensitivity loss at the 1-bit sampling with the efficiency $2/\pi \approx 0.63$[34].

The above-mentioned modulator ISL technology for the required bandwidth of $B_{max}$ does not exist and hence, the analogue amplitude modulation is being addressed in the Section 5. The widest channel bandwidths $B_i$ within $B_{max}$ are achieved using 1-bit ADC. Despite 2-bit sampling provides a better efficiency of circa 0.88 ($Q \approx 1.29$) which allows 1.4 times better sensitivity than the 1-bit sampling, the communication bandwidth restriction by $B_{max}$ disables any sensitivity benefit[34].

#### 4.2.1　System Noise Temperature at the Highest Frequency Channel

The main reflector antenna temperature depends mainly on its field of view, its aperture size and efficiency at the considered highest frequency channel of 637 GHz (Table 2), and its physical temperature.

The expected aperture efficiency of the EHIE main reflector is less than that of Planck, due to the bigger physical size of 3 m[30]. At this highest frequency of 637 GHz, the expected aperture efficiency is of 0.58 which corresponds to a half power beamwidth of 42.3 arcsec.

The Planck measurement of the flux in such beamwidth yields an antenna temperature of circa 5.87 K[①]. The temperatures measured by Planck towards several directions (Sgr A*, M87, 3C 66B) is up to 2.83 K, within a FoV of 10°. The latter temperature is adopted as the temperature outside the main lobe of the EHIE antenna radiation pattern.

The main reflector thermal shielding would allow its physical temperature to be near 120 K during the whole observation time of 6 months. The reflector radiation efficiency of 0.96 would then convert into an antenna noise temperature of 4.9 K. The noise temperature translated at the aperture of the main refle-

---

①http://pla.esac.esa.int/pla/#home

ctor is of circa 9.35 K.The assumed noise temperature of the LNA cooled down to a physical temperature of 55 K is 480 K. The expected receiver temperature is 5% higher which provides circa 504 K.#### 4.2.2 Effective Apertures

At this highest frequency of 637 GHz, the surface efficiency of the synthesized aperture is about 0.5 at the baseline localization accuracy of 0.11 mm (Section 2.1). Therefore, the effective area of the aperture $A_{\text{syn,e}} \approx 0.21 \times 10^{15} \text{m}^2$ is half its geometrical area of $\pi(D/2)^2$, at the absent tapering loss.

The effective area of each real aperture of the instrument is $A_{\text{real,e}} \approx 4.14 \text{m}^2$, at 637 GHz.

#### 4.2.3 Sensitivity and SNR Values

The obtained sensitivity is of circa $155 \times 10^6$ K, at a bandwidth of $B_2 \approx 2.22$ GHz and observation time of 6 months (Table 3). The shortening of the observation time reduces the sensitivity to circa $220 \times 10^6$ and $380 \times 10^6$ K for 3 and 1 months, respectively.

It is assumed for brevity that the incoming correlated flux of the Sgr A* does not depend on the baseline. The flux of the halo of the source model is of circa 5 mJy·beam$^{-1}$, at the synthesized antenna of $A_{\text{syn,e}}$ and the highest input frequency[35]. Further-study of Sgr A* flux source models could change the latter assumption significantly.

The SNR is of circa 4.9 at the halo (Table 3), acquiring both orthogonal polarizations. The flux peak of this Sgr A* model is of 43 mJy·beam$^{-1}$ and hence, the corresponding peak SNR is of circa 42.1. Assuming a noise temperature of the LNA of 300 K at the lower physical temperature of 15 K, the SNR in halo at 1-bit sampling is of 7.75, and the peak SNR is of 66.7.

Assuming a future modulator technology capable of exchanging 2-bit sampled signals, the SNR in halo and the peak are of 10.7 and 92, respectively.

These SNR values have been confirmed using the equation frequently used in radio astronomy[34]:

$$\sigma_{\text{noise}} = \frac{k(T_A + T_R)}{A_{\text{real,e}}\sqrt{2B_i \tau_{\text{obs}}}} \sqrt{Q}, \quad (8)$$

where $\sigma_{\text{noise}}$ is the standard deviation of the additive thermal noise at the correlator output and $k \approx 1.38 \times 10^{-23}$ J·K$^{-1}$ is the Boltzmann constant.

This interferometer sensitivity depends on: the baseline localization accuracy (which defines the $A_{\text{syn,e}}$), the real antenna losses (those define $A_{\text{real,e}}$ together with the launcher class), the receiver noise, and the telecommunication link capacity.

## 5 V-band ISLS

### 5.1 Link Budget

A single launch of both EHIE satellites on the SOYUZ launcher is proposed, which is compatible with an ISL antenna projected diameter of circa 0.8 m. Two single offset reflectors are needed to assure the aforementioned 100 dB isolation between the transmitted and the received signals. The feed horn antennas at different polarizations are placed inside the satellite body to decrease the height of the reflectors, launched at their final position. A single axis motor rotates the reflector. The rotation period is 4.5 hours, at these orbit radii (Table 1). The transmit and receive ISL antennas have to be placed on opposite sides of the spacecraft to avoid shadowing each other. Even at the shortest inter-satellite distance of 3.25 km (Table 1) and the highest centre frequency 68.5 GHz, the beamprint of a perfectly efficient antenna of diameter 0.8 m is of circa 17.2 m which is larger than the satellite dimensions.

The V-band ISLs are employed for both the derivation of frequencies of the local oscillators of the instrument and the exchange of the instrument signals as above-mentioned. As well, it is expected that future novel modem technology will replace the current analogue modulation by a digital one. The V-band ISL link budget aims an input signal-to-noise ratio of $\text{SNR}_{\text{in}} = 10$ dB, at the longest baseline of $D = 24\,712$ km (Table ). The Friis's transmission formula allows calculating the $\text{SNR}_{\text{in}}$ of 11 and 11.5 dB, at following input parameters.

(i) ISL antennas are 0.8 m dishes.

(ii) Antenna radiation and aperture efficiencies are 0.96 and 0.76, respectively.

(iii) Antenna noise temperature is 404 K accounting objects in main and minor lobes, antenna phy-

KUDRIASHOV Volodymyr et al.: System Design for the Event Horizon Imaging Experiment Using··· 261

sical temperature and efficiencies.

(iv) Output power of space-qualified Travelling Wave Tube Amplifier (TWTA) on the transmitter side is 75 W[36].

(v) Noise Figure (NF) of LNA is 2.8 dB. The NF of the space qualified LNA is 2.5 dB. This LNA can to be tuned to 59.4~64.2 GHz band. A newly-designed and batch qualified LNA is aimed, at the second V-band ISL band of 66.2~70 GHz.

(vi) Receiver noise temperature is 5% higher than the LNA noise temperature, as in Subsection 4.2.1.

(vii) At $r_T = 20$ km, the absence of any atmospheric effects allows the free space transmission loss below 216.1 and 217.0 dB for centre wavelengths of 4.9 and 4.4 mm, respectively. Antenna de-pointing allows keeping the receiver input signal within its dynamic range, on short baselines.

(viii) Minor losses of both antenna pointing motor[37] and the aforementioned star sensor, flanges of V-band feeding lines, and frequency shift due to inter-satellite velocity are below 0.2 dB in total.

These $SNR_{in}$ margins of 1 and 1.5 dB may mitigate requirements to hardware or increase the link performance.

The upper bound for the Additive White Gaussian point-to-point communication channel capacity set by the Shannon-Hartley theorem is of circa $16.5 \times 10^9$ bit·s$^{-1}$, at the $SNR_{in} = 10$ dB of the V-band ISL and the bandwidth $B_{max}$. The implementation of such high-speed digital modulator and demodulator as a single carrier ultra-wideband system is considering technically challenging based on the existing digital technologies. Other solutions based on multi-carrier modulator and demodulator implementation may exist. However, such solutions are expected to lead to excessive power and mass budget. As an alternative, the analogue frequency conversion and transmission is addressed.

### 5.2　Cross-link Leakage

The ratio of received and transmitted power is calculated using the Friis's transmission formula, for an antenna diameter of 0.8 m and the above-mentioned parameters. These ratios are of $-110.5$ and $-109.6$ dB for centre wavelengths of 4.9 and 4.4 mm, respectively. Filters at both receiver and transmitter can assure an isolation of circa 100 dB between two non-overlapping frequency bands (Subsection 3.1, Fig. 3). The additional inter-channel isolation of 50±10 dB is obtained by the use of different antennas for transmission and reception, and by the operation at different circular polarizations[38]. Thus, the ratio of received to transmitted leakage power is of 39±10 dB.

### 5.3　Accommodation of the ISL Antennas

To avoid mutual blockage, the ISL antennas are mounted on opposite sides. The other 4 sides of the platform would be used for: the main reflector, the solar array, propulsion and the interface to the launcher[38]. The rotation axis of the ISL antennas is perpendicular to the orbital plane. The 1-axis motors are redundant, as well as both transmission and reception channels.

### 5.4　Range Rate Measurement Concept

The dual one-way range rate measurement concept allows cancellation of clock errors[39-43]. In order to obtain the local oscillator component $f_1$ (and $f_2$), the master clock frequency $f_0 = 10$ MHz is multiplied by $k_1 = 5940$ (and $k_2 = 6613$). The master clock frequency error $\varepsilon_1$ (and $\varepsilon_2$) is fed to the multiplier too (Fig. 5). The transmitted signal frequency is expressed as: $f_1 = k_1(f_0 + \varepsilon_1)$. The signal at $f_1$ (and the other satellite signal at $f_2 = k_2(f0 + \varepsilon_2)$) is transmitted towards the other satellite. The signal experiences a Doppler frequency shift $f_{D_1}$ (and $f_{D_2}$) due to the relative velocity between satellites. The frequency of the incoming signal $f'_1 = f_1 + f_{D1}$ (and $f'_2 = f_2 + f_{D2}$) is down-converted using the signal at $f_2$ (and $f_1$) as local oscillator. Each down-converted frequency is measured. At a time instance $i$, the instantaneous clock errors and frequency shifts allow expressing the measured frequencies as:

$$f_{M_1,i} = f_{D_1,i} + [f_0(k_1 - k_2) + k_1\varepsilon_{1,i} + k_2\varepsilon_{2,i}],$$
$$f_{M_2,i} = f_{D_2,i} - [f_0(k_1 - k_2) + k_1\varepsilon_{1,i} + k_2\varepsilon_{2,i}]. \quad (9)$$

The sum of the measured frequencies $f_{M_1,i} + f_{M_2,i}$ provides the sum of the Doppler frequency shifts

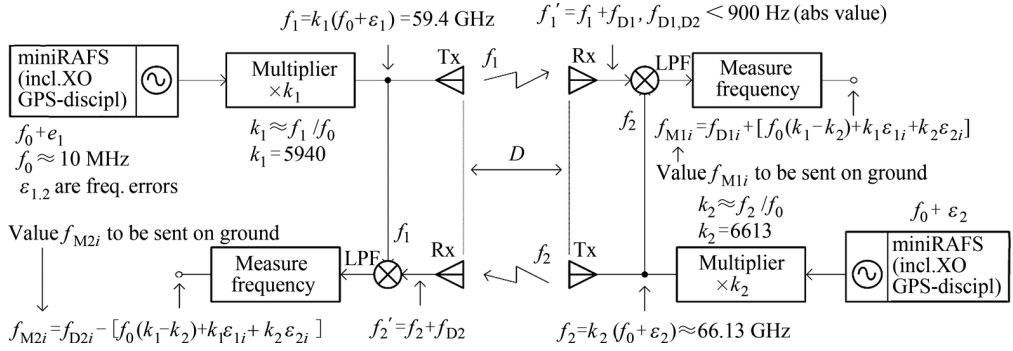

Fig. 5  Block diagram of V-band range-rate measurement which allows the clock error cancellation

$f_{D_{1,i}} + f_{D_{2,i}}$, i.e. cancels out errors of clock sources. The adding is to be done on the ground. Hence, the measured frequency values $f_{M_{1,i}}$ and $f_{M_{2,i}}$ must are time stamped and sent to ground.

### 5.5 Fundamental Accuracy of Range-rate Measurement

The lower bound of the range rate measurement accuracy depends on the clock errors and the SNR. The lower bound of the variance of the unbiased estimator (the Cramer-Rao Lower Bound) establishes the fundamental accuracy of the radial velocity measurement as[44]:

$$\sigma_{vr1,2} = \frac{\lambda_{1,2}}{\tau q}, \qquad (10)$$

where $\sigma_{vr1,2}$ are the accuracies at wavelengths $\lambda_{1,2}$ corresponding to the frequencies $f_1$ and $f_2$, $\tau$ is the integration time, and $q = \sqrt{2 \times 10^{SNR_{out}/10}}$ is the quality factor defined by the output SNR, which equals the product of the $SNR_{in}$ and the signal processing gain (the time-bandwidth product). Assuming both the absence of frequency error and exact knowledge of $\tau$, then the one way range rate fundamental measurement accuracy is expressed as:

$$\sigma_{rr1,2} = \tau \sigma_{vr1,2}, \qquad (11)$$

Assuming an ideal frequency multipliers and an accuracy of the master clock frequency $\sigma_{cl} = 2 \times 10^{-10}$ the accuracies of frequencies $f_1$ and $f_2$ can be estimated as $\sigma_{f_{1,2}} = k_{1,2} f_0 \sigma_{cl}$ that yield 11.9 and 13.2 Hz, respectively①. The radial velocity between satellites changes along the observation time and hence, the allowed integration time is to be refined. For input bandwidths of $\pm \sigma_{f1,2}$ and an integration time of 180 s (and 1 s), the time-bandwidth products for $f_1$ and $f_2$ are of circa 36.3 and 36.8 dB (13.8 and 14.2 dB), respectively. The $SNR_{out}$ values are 10 dB higher than the latter, as $SNR_{in} = 10$ dB.

For an integration time of 180 s (1 s), the calculated $\sigma_{vr1,2}$ are of circa 96 and 81.4 nm·s$^{-1}$ (230.6 and 197.8 $\mu$m·s$^{-1}$). The calculated $\sigma_{rr1,2}$ are of circa 17.3 and 14.7 $\mu$m (0.73 and 0.63 mm). The latter are more than 4 times better than the above-mentioned requirement for $\sigma$ (Fig. 1).

For signals with non-random initial phase, the fundamental accuracy is $2\pi$ times better than for signals with a random initial phase[45]:

$$\sigma'_{rr1,2} = \frac{\lambda_{1,2}}{2\pi q}. \qquad (12)$$

The use of an IQ receiver validates the non-random initial phase assumption. Thus, for an integration time of 180 s (1 s) the improved values of the range rate fundamental accuracy are of circa 2.7 and 2.3 $\mu$m (36.7 and 31.5 $\mu$m).

## 6  K-band ISLS

### 6.1  Link Budget

Accounting for the performance degradation at the end of the lifetime, the expected noise figure of the whole receiver is of 2.5 dB, at 27.5~31 GHz frequency band②. At the transmitter, the TWTA output power

---

① http://www.spectratime.com/uploads/documents/ispace/iSpace_miniRAFS_Spec.pdf
② http://www.ommic.fr/download/CGY2260UHC1_PDS_170217.pdf
　http://www.lucix.com/images/pdfs/2012_brochure_LNAs_v.b.pdf

is of 45 W, covering a bandwidth up to 1 GHz[①,[46]]. K-band ISLs benefit from both the absence of atmospheric effects, for $r_T = 20$ km, and from the V-band reflectors. The sources of minor losses are same to aforementioned ones. The Friis's transmission formula allows calculating the achievable $SNR_{in}$ of 10.7 and 11.2 dB, at the longest baseline.

### 6.2　Range and Range-rate Measurement Concept

The K-band ranging benefits of a bandwidth of $B_{K,max} = 578.8$ MHz. The dual one-way ranging code is stored in a memory. The code being converted into analogue form is modulated by the telemetry information which is converted into analogue form too. The output of the modulator is fed to the frequency up-converter. The local oscillator of the up-converter is the signal of the master clock at the output of the frequency multiplier. The band-pass filter and the TWTA assure the concordance with the radio regulations and the inter-channel isolation. At the receiver side, a super-heterodyne receiver and demodulator allow to feed the incoming ranging code to the ADC and the cross-correlator. The code memory at the receiver side delivers the reference signal for the cross-correlator dedicated to measuring both the range and the range rate.

### 6.3　Cross-link Leakage

The transmit-receive isolation of the K-band links depends on both the input ratio of the transmitted and the received signal power and the signal processing gain. The value of the former (calculated using the Friis's transmission formula) is of circa $-118$ dB. Both the above-mentioned filter and the operation at different polarizations assure the expected inter-channel isolation of $150\pm10$ dB.

### 6.4　Range and Velocity Fundamental Accuracies

The integration time for the K-band inter-satellite ranging is limited by the long term accuracy of the master clock $\sigma_{cl} = 2\times10^{-10}$. The calculated accuracy of the sum of uncorrelated frequency errors of two centre frequencies of the ranging signals is 7.5 Hz. At the $\pm3$ sigma interval of the latter accuracy of circa 44.9 Hz, the integration time at Doppler measurement is shorter than 22.3 ms. At this integration time and the $B_{K,max} = 578.8$ MHz, the time-bandwidth product is of circa 71 dB and hence, the output SNR is of circa 81 dB. The mean of the centre frequencies is 26.4 GHz. At this frequency and an integration time of 22.3 ms, the accuracy of the radial velocity measurement is of circa 31.7 $\mu$m·s$^{-1}$. Assuming the absence of a clock error component in the measurement error, as for the V-band above, the fundamental accuracy of the range rate error is 0.7 $\mu$m.

The range resolution $\delta R$ is of circa 0.52 m, at the absent tapering of the spectra of signals. The fundamental accuracy of the range measurement is calculated as Eq. (10) in which the $\lambda_{1,2}/\tau$ has been substituted by the $\delta R$[44]. The calculated accuracy is of circa 32.3 $\mu$m. Both values are much less than the wavelength of the GPS L1 C/A signal.

## 7　Conclusions

This paper addressed the frequency plan, the sensitivity, and the sub-systems of the space-to-space interferometry concept from PECMEO applicable to the imaging of Sgr A*. The essence of the concept is the reduction of the main reflector size by the substitution of the conventional technique for precise localization of the interferometer baseline by the one based on GNSS and AODS.

The paper proposes requirements for the localization knowledge of the interferometer baseline vector. It also provides information about the GNSS availability and the output data from the attitude and orbit determination systems.

The paper proposes a frequency plan for the instrument and its inter-satellite links compatible with international radio regulations. The concept for the measurement of range and range rate through ISLs as well as the data exchange over the ISLs are presented. The interferometer block diagram is described. The sensitivity is estimated.

---

① https://www.thalesgroup.com/sites/default/files/asset/document/thales_space_k_ka_band.pdf
　http://www.necnets.co.jp/en/products/twt/seihin/ld7314.html

The power budget for both inter-satellite links is given as well as their critical components. The calculated fundamental range measurement accuracies are given as well as the range resolution.

The paper results are applicable to the whole PECMEO system. The estimated mass per satellite is of circa 1 ton. The budget estimate is within that of a middle-class science mission of the European Space Agency.

**Acknowledgements** The authors are grateful to the following people from ESA: to B Duesmann who contributed orbit selection, to S Bras for review and suggestion on the inertial navigation sensors, to D de Wilde who designed the supporting plate for the purpose of twin satellites and confirmed the possibility of accommodation in the fairing of the Soyuz, to P Piironen, V Tuomas Kangas, N Ayllon and H Barre who advised on microwave equipment, to P de Maagt, M Van der Vorst and Salghetti Drioli L, J Tauber for their support on the parameters of antennas; from ICE-IEEC to W Li who simulated the GNSS visibility at the initial configuration of 2 navigation antennas; from Radboud University to M Moscibrodzka and to C Brinkerink, F Roelofs, H Falcke for the suggestion on the flux of the source, the sensitivity calculation, and the support of the follow-up work; *from LNDES, IRE NASU* to K Lukin for his encouragement and interest in the PECMEO; *from JIVE* to L Gurvits for his remarks and recommendations. Authors are grateful to unknown referees for their valuable comments.